\documentclass[aps,amsmath,twocolumn,a4paper,blue]{revtex4}

\usepackage{graphicx}
\usepackage[dvips]{color}
\begin{document}

\title{Single-File Diffusion in a Box}

 \author{L.~Lizana}
 \email{lizana@fy.chalmers.se}
 \affiliation{Department of Chemical and Biological Engineering,
              Chalmers University of Technology, Gothenburg, Sweden}

 \author{T.~Ambj\"ornsson}
 \email{ambjorn@mit.edu}
 \affiliation{Department of Chemistry, Massachusetts
              Institute of Technology, Cambridge, MA 02139}

\date{\today}

\begin{abstract}
We study diffusion of (fluorescently) tagged hard-core interacting
particles of finite size in a finite one-dimensional system. We find
an exact analytical expression for the tagged particle probability
density using a Bethe-ansatz, from which the mean square
displacement is calculated. The analysis show the existence of three
regimes of drastically different behavior for short, intermediate and
large times. The results show excellent agreement with stochastic
simulations (Gillespie algorithm). 
\end{abstract}

\maketitle

%
%


{\it Introduction. - } Recent advances in the manufacturing of
nanofluidic devices allow studies of geometrically constrained nano -
sized particles in quasi one-dimensional systems in which crowding and
exclusion effects are important \cite{KKEDJJO,CD}. Situations
where large molecules are hindered to overtake also occurs in living
systems as, for instance, protein diffusion along DNA
\cite{ABKM}. Furthermore, biological cells are characterized by a high
degree of molecular crowding \cite{Ellis}.

\begin{figure}
 \includegraphics[width = 0.7\columnwidth]{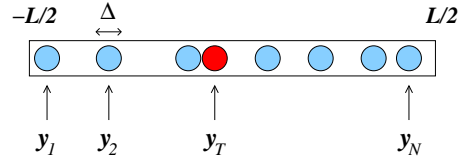}
 \caption{(Color online) Diffusing particles where mutual passage is
  excluded, $i.e$ $y_j\leq y_{j+1}-\Delta$ for $j=1,\ldots,N-1$.}
 \label{fig:ParticlesInBox}
\end{figure}

In this Letter, with experiments in mind, akin to \cite{LMCDR}, we
focus on diffusive motion of tagged finite-sized hard-core interacting
particles (unable to overtake) (Fig.  \ref{fig:ParticlesInBox}). Such
single-file systems show interesting behavior where the $t^{1/2}$ -
scaling ($t$ denotes time) of the mean square displacement ${\cal
S}(t) =\langle [y_{\cal T}(t) -y_{{\cal T},0}]^2\rangle$ in position
$y_{\cal T}(t)$ of the tagged particle $[y_{{\cal T},0} \equiv y_{\cal
T}(0)]$, for an infinite system with fixed concentration, is most
striking ($\langle . \rangle$ denotes ensemble average). Also, the
probability density function (PDF) $\rho(y_{\cal T},t|y_{{\cal
T},0})\equiv\rho_{\cal T}$ is Gaussian \cite{MK,TEH,JALA}. Even though
single-file diffusion has received much attention
\cite{SFD_THE,RKH,GMS,JALA,SFD_EXP}, to our knowledge very few exact
results are given for finite-sized particles in finite systems
\cite{FEAL}. One exception is \cite{RKH} where the PDF for $N$
diffusing point particles on a finite interval was obtained. However,
asymptotic expressions were only given when the system was made
infinite (keeping the concentration finite). Here, we go beyond
previous results in the following ways. First, finite-sized particles
are considered and we show that the $N$-particle PDF can be written as
a Bethe-ansatz solution. Second, we perform a (non-standard) large
$N$-analysis of $\rho_{\cal T}$, keeping the system size {\em finite},
showing, for the first time, the existence of three dynamical regimes:
$(i)$ $t \ll \tau_{\rm coll}=1/\varrho^2D$ where $\tau_{\rm coll}$
denotes mean collision time, $D$ the diffusion constant, and
$\varrho=N/L$ particle concentration where $L$ is the length of the
system, $(ii)$ $\tau_{\rm coll} \ll t \ll \tau_{\rm eq}$ where
$\tau_{\rm eq}=L^2/D$ is the equilibrium time, and $(iii)$ $t\gg
\tau_{\rm eq}$. Notably, only $(i)$ and $(ii)$ are found in infinite
systems. Asymptotic expressions for $\rho_{\cal T}$ are derived in
regimes $(i)$-$(iii)$ and good agreement with (stochastic) Gillespie
simulations \cite{Gillespie},
is demonstrated.

%
%

%
%

{\it Statement of the problem. - } We study $N$ hard-core interacting
particles with linear size $\Delta$ diffusing in a one-dimensional box
of length $L$ (Fig. \ref{fig:ParticlesInBox}).  The probability of
finding the particles at positions $\vec{y}= (y_1,\ldots,y_N)$ at time
$t$, given that they initially were at $\vec{y}_0=(y_{1,0},\ldots,y_{N,0})$,
is contained in the $N$-particle conditional PDF ${\cal P}
(\vec{y},t|\vec{y}_0)$ which is governed by the diffusion equation
\begin{equation}\label{eq:DiffusionEq}
 \frac{ \partial {\cal P} (\vec{y},t|\vec{y}_0)}{\partial t}
  = D \left(
    \frac{\partial^2}{\partial y_1^2} + \ldots +
    \frac{\partial^2}{\partial y_N^2}
      \right)
    {\cal P}(\vec{y},t|\vec{y}_0).
\end{equation}
Neighboring particles $j$ and $j+1$ are unable to overtake
\begin{equation}\label{eq:Condition_A}
 D \left. \left(
      \frac{\partial}{\partial y_{j+1}}-\frac{\partial}{\partial y_j}
   \right)
   {\cal P}(\vec{y},t|\vec{y}_0)\right|_{y_{j+1}-y_j=\Delta} = 0,
\end{equation}
ensuring that $y_{j+1}-y_j\geq \Delta$ $\forall t$ provided that
$y_{j,0} < y_{j+1,0}-\Delta$. The boundaries at $\pm L/2$ are
reflecting
\begin{equation}\label{eq:Condition_B}
 D \left. \frac{{\cal P}(\vec{y},t|\vec{y}_0)}
               {\partial y_1}\right|_{y_1=-(L-\Delta)/2}
 \hspace{-0.25cm} =
 D \left. \frac{{\cal P}(\vec{y},t|\vec{y}_0)}
                {\partial y_N}\right|_{y_N=(L-\Delta)/2}
     \hspace{-0.25cm} = 0,
\end{equation}
and the initial PDF is given by
\begin{equation}\label{eq:Condition_Init}
 {\cal P}(\vec{y},0|\vec{y}_0) =
  \delta(y_1-y_{1,0}) \cdots
  \delta(y_N-y_{N,0}),
\end{equation}
where $\delta(z)$ is the Dirac delta function. The tagged particle
PDF studied here is given by \cite{AMLI}
\begin{eqnarray}\label{eq:PDF_tagged1}
 \rho_{\cal T}&=&
 \frac{1}{\rho_{\rm eq,{\cal T}0}}
 \int_{\cal R} dy_1'\cdots dy_N'
 \int_{{\cal R}_0} dy_{1,0}'\cdots dy_{N,0}' \nonumber \\
&&  \times \,
    \delta(y_{\cal T}-y_{\cal T}') \,
    \delta(y_{{\cal T},0}-y_{{\cal T},0}') \,
    {\cal P}(\vec{y}^{\,_{'}}\hspace{-0.15 cm},t|\vec{y}_0^{\,_{'}}) \,
    {\cal P}_{\rm eq}(\vec{y}_0^{\,_{'}}) \ \ \ \ \
\end{eqnarray}
where
$ \rho_{\rm eq,{\cal T}0} =
 \int_{{\cal R}_0} dy_{1,0}'\cdots dy_{N,0}'
 \delta(y_{{\cal T},0}-y_{{\cal T},0}')
 {\cal P}_{\rm eq}(\vec{y}_0^{\,_{'}})$,
with integration regions
${\cal R} = \{ y_{j+1}-y_j \geq \Delta, j = 1,\ldots,N-1;
y_1\geq-(L -\Delta)/2;
  y_N \leq (L-\Delta)/2  \}$
and
${\cal R}_0 = \{ y_{j+1,0}-y_{j,0} \geq \Delta, j = 1,\ldots,N-1;
 y_{1,0}\geq -(L -\Delta)/2;
   y_{N,0} \leq (L-\Delta)/2 \}$.
Initially, the particles are distributed according to the
equilibrium density,
\begin{equation} \label{eq:PDF_eq1}
 {\cal P}_{\rm eq}(\vec{y}) = \frac{N!}{L^{N}}
  \Pi_{l=1}^{N-1}\theta(y_{l+1}-y_{l} - \Delta),
\end{equation}
$i.e$ the particles are distributed uniformly in the box. The function
$\theta(z)$ is the Heaviside step function.

%
%

{\it Bethe - ansatz solution. - } The (coordinate) Bethe-ansatz yields
an integral representation of the PDF in momentum space
\cite{GMS}. The Bethe ansatz satisfying Eqs.~(\ref{eq:DiffusionEq}) -
(\ref{eq:Condition_Init}) is
 \begin{eqnarray}\label{eq:BetheAnsatz}
  {\cal P} (\vec{x},t|\vec{x}_0)
 & = &  \int_{-\infty}^\infty\cdots\int_{-\infty}^\infty
    \frac{dk_1\cdots dk_N}{(2\pi)^N} \,
    e^{-E(\vec{k}) t} \phi(k_1,x_{1,0}) \nonumber \\
&& \hspace{-1cm}
  \times \ \cdots \phi(k_N,x_{N,0})
  \big[ \
     e^{i(k_1x_1 + k_2x_2 + k_3x_3 + \ldots + k_Nx_N)}
      \nonumber \\
 && \hspace{-1cm}
     + \ S_{21} e^{i(k_2x_1 + k_1x_2 + k_3x_3\ldots + k_Nx_N)}
     \nonumber \\
 && \hspace{-1cm}
     + \ S_{21}S_{31} e^{i(k_2x_1 + k_3x_2 +  k_1x_3+ \ldots + k_Nx_N)}+
     \ldots
    \big],
 \end{eqnarray}
where $x_1,\ldots,x_N$ and $x_{1,0},\ldots,x_{N,0}$ are given in terms
 of $\vec{y}$ and $\vec{y}_{0}$ according to ($j=1,\ldots,N$)
\begin{equation}\label{eq:finite_transfm}
 \ell = L - N\Delta, \ \ \
 x_{j(,0)} = y_{j(,0)} - \Delta \left(j-\frac{N+1}{2} \right),
\end{equation}
where
$-\ell/2 \leq x_1 \leq x_2 \leq \ldots \leq x_N \leq \ell/2$
and
$-\ell/2 \leq x_{1,0} \leq \ldots \leq x_{N,0} \leq \ell/2$.
In fact, transformation (\ref{eq:finite_transfm}) effectively maps
Eqs.~(\ref{eq:DiffusionEq})-(\ref{eq:Condition_Init}) onto a
point-particle problem. The bracket in Eq.~(\ref{eq:BetheAnsatz})
contains $N!$ terms corresponding to all permutations of momenta
$\vec{k}=k_1,\ldots, k_N$. The quantities $S_{lj}$ are scattering
coefficients which contain information about the pair interaction
between particles $l$ and $j$, and are in general functions of $k_l$
and $k_j$. For the case of a pair interaction on the form given by
Eq.~(\ref{eq:Condition_A}), $S_{lj}\equiv 1$ ($S_{lj}\equiv 0$ for
non-interacting particles) \cite{AMLI}. The time dependence enters
through $e^{-E(\vec{k})t}$ with dispersion relation ("energy")
$E(\vec{k})= D\,(k_1^2+\ldots+k_N^2)$, obtained from
Eq. (\ref{eq:DiffusionEq}).  The functions $\phi(k_j,x_{j,0})$ carry
information about the boundary and initial conditions
Eqs.~(\ref{eq:Condition_B})-(\ref{eq:Condition_Init}), and for the
finite box studied here
$\phi(k_j,x_{j,0}) = 2\sum_{m=-\infty}^\infty
                \cos[k_j(x_{j,0}+\ell/2)]\,
                e^{ik_j(2m+1/2)\ell}$,
which was found using the method of images \cite{AMLI}. For an
infinite system ($\ell \rightarrow \infty$), $\phi(k_j,x_{j,0}) =
e^{-ik_j x_{j,0}}$ \cite{GMS}.

Performing integrations over $k_1,\ldots, k_N$,
Eq.~(\ref{eq:BetheAnsatz}) is rewritten as
\begin{eqnarray}\label{eq:ReflectionPrinciple}
   {\cal P} (\vec{x},t|\vec{x}_0) &=&
   \psi(x_1,x_{1,0};t) \, \psi(x_2,x_{2,0};t) \cdots \psi(x_N,x_{N,0};t)
   \nonumber\\
  &&\hspace{-1cm}  + \
   \psi(x_1,x_{2,0};t) \, \psi(x_2,x_{1,0};t) \cdots \psi(x_{N,0},k_N;t)
   \nonumber\\
  &&\hspace{-1cm} + \
    {\rm remaining \, permutations} \, {\rm of}  \, x_{1,0},\ldots,x_{N,0},
\end{eqnarray}
where
$\psi(x_j,x_{l,0};t)  =
 \int_{-\infty}^\infty \frac{dk_j}{2\pi}
  \phi(k_j,x_{l,0}) e^{ik_j x_j} e^{-Dk_j^2t}$
is the integral representation (inverse Fourier transform) of the
(free) single particle PDF for particle $j$. Notably, as $\Delta
\rightarrow 0$, the $N$-particle PDF in \cite{RKH} is recovered. The
single particle PDF is also found in \cite{RKH} (for $\Delta=0$) but
in an unsuitable form for studies of finite systems. We obtained a
more convenient expression, in which the large time limit is easily
tractable, by finding the Laplace transform to $\psi(x_j,x_{l0};t)$,
calculating the poles, and inverting it back using the corresponding
residues enclosed within the Bromwich contour~\cite{AMLI}:
\begin{eqnarray}\label{eq:FreeParticleInBox}
 \psi(x_j,x_{l,0};t) &=& \frac{1}{\ell}\Big\{ 1
 + \sum_{m=1}^{\infty}F_m(t) \nonumber \\
 &\times& \Big[
 \nu_m^{(+)} \cos \left(\frac{\pi m x_j}{\ell}\right)
             \cos \left(\frac{\pi m x_{l,0}}{\ell}\right) \nonumber \\
 &&+\nu_m^{(-)} \sin \left(\frac{\pi m x_j}{\ell}\right)
                \sin \left(\frac{\pi m x_{l,0}}{\ell}\right)
 \Big] \Big\}, \ \ \ \ \
\end{eqnarray}
where $\nu_m^{(\pm)}=1\pm(-1)^m$ and $F_m(t) =
e^{-(m\pi)^2\frac{Dt}{\ell^2}}$.

%
%
%

\begin{figure}
 \includegraphics[width = \columnwidth]{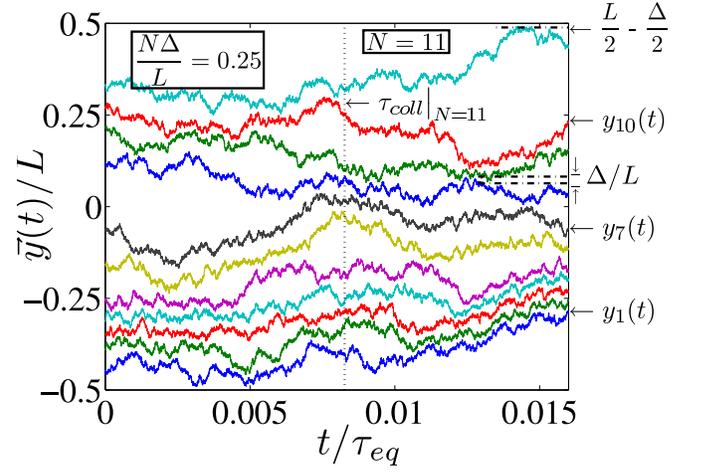}
 \caption{(Color online) Particle trajectories, generated by the
   Gillespie algorithm, in a system where $N=11$ .}
 \label{fig:Gill_traj}
\end{figure}


%
%

{\it Tagged particle density. - } Integrating
Eq.~(\ref{eq:ReflectionPrinciple}) according to
Eq.~(\ref{eq:PDF_tagged1})~\cite{RKH} leads to an exact form of the
tagged particle PDF in terms of Jacobi polynomials
$P_n^{(\alpha,\beta)}(z)$ \cite{ABST}, given by
\footnote{For $\Delta=0$, Eq.~(\ref{eq:PDF_tagged2}) is also found in
  \cite{RKH}[Eq. (61)] and is related to Eq. (\ref{eq:PDF_tagged2})
  through elementary relations of the Jacobi Polynomial, found in
  $e.g.$ \cite{ABST}.}
\begin{eqnarray}\label{eq:PDF_tagged2}
\rho_{\cal T} &=& \frac{(N_R + N_L - 1)!}{N_L!N_R!}
           (\psi_L^L)^{N_L}(\psi_R^R)^{N_R}          \nonumber \\
            && \hspace{-0.7cm} \times \Big\{
            (N_L+N_R)\psi \, \Phi(0,0,0;\xi)
           + N_L^2 \frac{\psi_L\psi^L}{\psi_L^L} \Phi(1,0,0;\xi)
                                                     \nonumber \\
     &&  \hspace{-0.4cm}
     +\, N_R^2 \frac{\psi_R\psi^R}{\psi_R^R}
                                            \Phi(0,1,0;\xi)
                                                     \nonumber \\
    && \hspace{-0.4cm}
              +\, N_RN_L
                  \left[
                     \frac{\psi_R\psi^L}{\psi_R^L} +
                 \frac{\psi_L\psi^R}{\psi_L^R}
                 \right]\Phi(0,0,1;\xi)
                                          \Big\}.
\end{eqnarray}
where
$\xi= \xi(y_{\cal T},y_{{\cal T}, 0},t) = \psi_R^L\psi_L^R / (\psi_R^R\psi_L^L)$,
$N_L (N_R)$ denotes the number of neighbors to the left (right) of the
tagged particle, and
\begin{eqnarray}\label{eq:JacobiPol}
  \Phi(a,b,c;\xi)&=&
  \frac{(N_L-(a+c))!(N_R-b)!}{(N_L+N_R-(a+b+c))!}
  (1-\xi)^{N_L-(a+c)}   \nonumber\\
  &&\times  \
 \xi^c \,
  P_{N_L-(a+c)}^{(c,N_R-N_L+a-b)}\left(\frac{1+\xi}{1-\xi}\right).
\end{eqnarray}
Arguments $y_{\cal T}$, $y_{{\cal T}, 0}$ and $t$ were left implicit. Also,
\begin{eqnarray}\label{eq:psi_int}
 \psi_L^L &=& \frac{1}{2}+\frac{x_{\cal T}}{\ell}+
   \left( \frac{1}{2}+\frac{x_{{\cal T},0}}{\ell} \right)^{-1}
  \sum_{m=1}^\infty  I_m F_m(t)\nonumber \\
 \psi_R^R &=& \frac{1}{2}-\frac{x_{\cal T}}{\ell}+
   \left( \frac{1}{2}-\frac{x_{{\cal T},0}}{\ell} \right)^{-1}
  \sum_{m=1}^\infty  I_m F_m(t)\nonumber \\
 \psi^L & = &\frac{1}{2}+\frac{x_{\cal T}}{\ell} +
       \sum_{m=1}^\infty J_m (x_{\cal T},x_{{\cal T},0}) F_m(t) \\
 \psi_L &=& \frac{1}{\ell} +  \frac{1}{\ell}
         \left(
                \frac{1}{2}+\frac{x_{{\cal T},0}}{\ell}
        \right)^{-1}
        \sum_{m=1}^\infty J_m (x_{{\cal T},0},x_{\cal T}) F_m(t)\nonumber\\
 \psi_R &=& \frac{1}{\ell} -  \frac{1}{\ell}
         \left(
                \frac{1}{2}-\frac{x_{{\cal T},0}}{\ell}
        \right)^{-1}
            \sum_{m=1}^\infty J_m (x_{{\cal T},0},x_{\cal T}) F_m(t)\nonumber
 \end{eqnarray}
where
\begin{eqnarray}\label{eq:psi_int2}
 I_m & = &\frac{1}{(m\pi)^2}
 \Big[ \nu_m^{(+)} \sin \left(\frac{m\pi x_{\cal T}}{\ell}\right)
            \sin \left(\frac{m\pi x_{{\cal T},0}}{\ell}\right) \nonumber\\
  &&+ \,\nu_m^{(-)} \cos \left(\frac{m\pi x_{\cal T}}{\ell} \right)
             \cos \left(\frac{m\pi x_{{\cal T},0}}{\ell}\right) \Big]
               \nonumber \\
 J_m(z,z') & = &\frac{1}{m\pi}
 \Big[ \nu_m^{(+)} \sin \left(\frac{m\pi z }{\ell}\right)
           \cos \left(\frac{m\pi z' }{\ell}\right) \nonumber\\
  &&- \,\nu_m^{(-)} \cos  \left(\frac{m\pi z }{\ell}\right)
               \sin  \left(\frac{m\pi z' }{\ell}\right)\Big]
\end{eqnarray}
Normalization gives $\psi^R=1-\psi^L$, $\psi_L^R=1-\psi_L^L$ and
$\psi_R^L=1-\psi_R^R$, which completely determines $\rho_{\cal T}$. A
\texttt{MATLAB} implementation of $\rho_{\cal T}$ using Eqs.
(\ref{eq:PDF_tagged2})-(\ref{eq:psi_int2}) is available upon request.

%
%
%
%

\begin{figure}
 \includegraphics[width = 0.9\columnwidth]{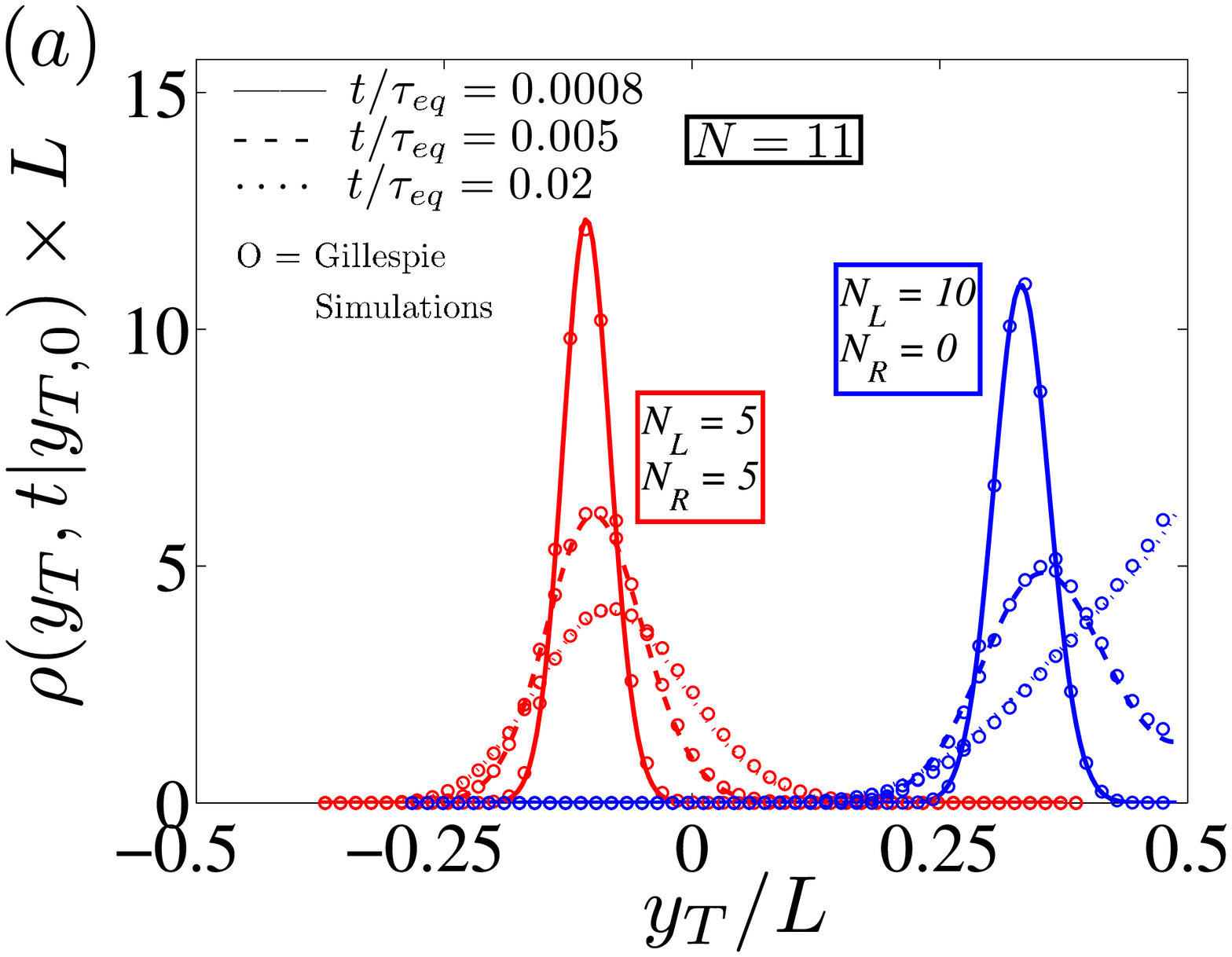}
 \includegraphics[width = 0.9\columnwidth]{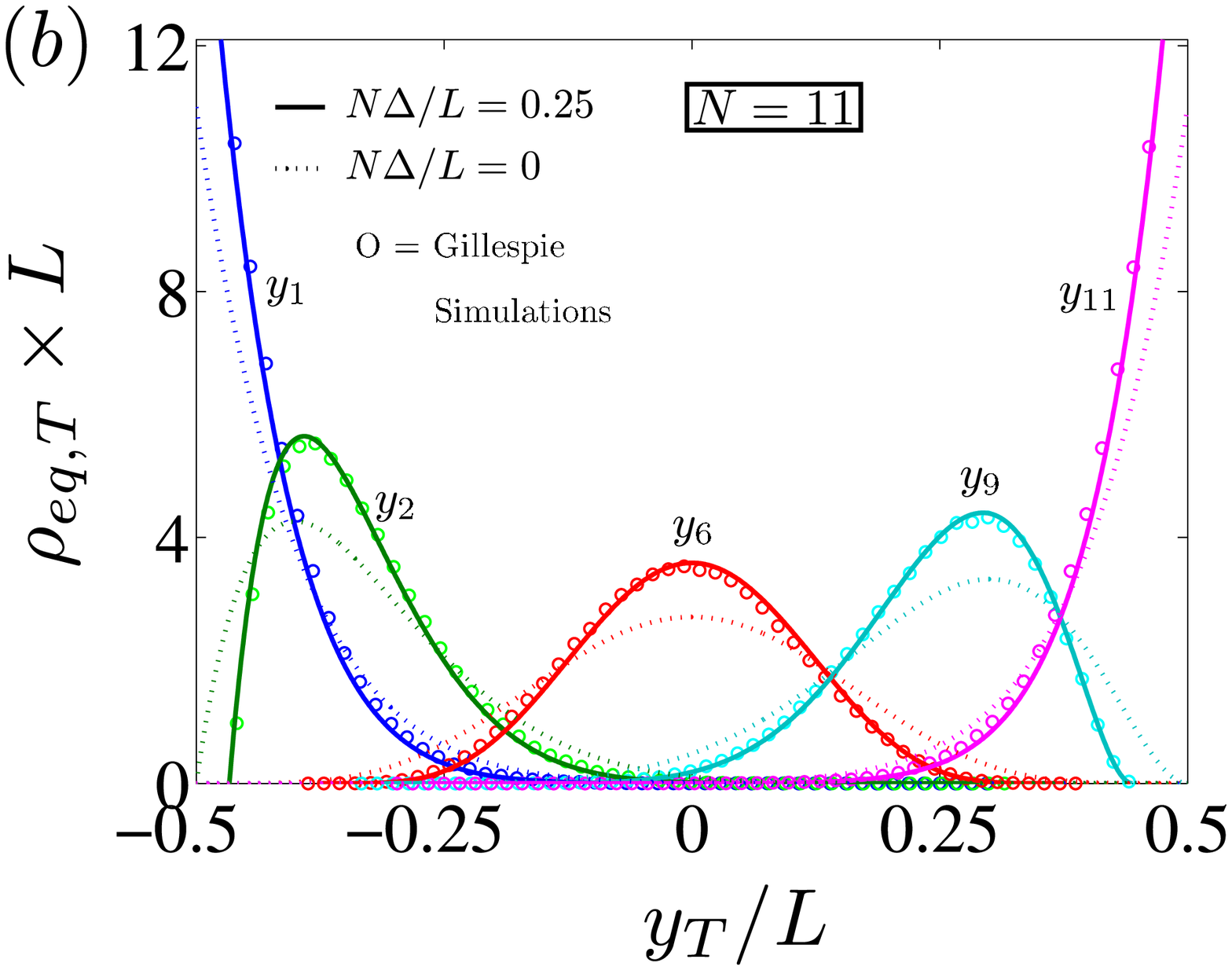}
 \caption{(color online) $(a)$ Tagged PDF [Eq.~(\ref{eq:PDF_tagged2})]
 depicted for the middle (red) and the rightmost (blue) particle,
 where $N\Delta/L = 0.25$, at three instants of time. The initial
 conditions were $y_{5,0}/L=-0.15$ and $y_{11,0}/L=0.3$.
 $(b)$ Equilibrium density [Eq.~(\ref{eq:PDF_eq2})] compared to the
 case of point-particles and a stochastic simulation ($t/\tau_{\rm eq}=2$)
 denoted by ($\circ$). The agreement between the simulations and the
 analytical results was checked using a $\chi^2$-test with
 significance level $\alpha = 0.01$.
 In the simulations, 500 lattice points and $10^5$ ensembles were~used.}
 \label{fig:Densty_vs_t}
\end{figure}
%

%
Figures \ref{fig:Gill_traj} and \ref{fig:Densty_vs_t} illustrate the
typical behavior of the finite single-file system via stochastic
simulations and $\rho_{\cal T}$. Figure \ref{fig:Gill_traj} shows
particle trajectories produced by the Gillespie algorithm (a Monte
Carlo-like algorithm based on a lattice model which is equivalent to
the master equation \cite{Gillespie}).  Figure \ref{fig:Densty_vs_t}
(a) illustrates the time evolution of $\rho_{\cal T}$ for one tagged
particle in the middle of the ensemble, and one by the edge.
Snapshots of the PDFs are given at short (solid), intermediate
(dashed) and large times (dotted). Notice the excellent agreement
between the analytical result Eq.~(\ref{eq:PDF_tagged2}) and the
stochastic simulation.
Panel (b) contains examples of the equilibrium PDF compared to the
point-particle case $\Delta=0$.

%
%

%
%
%
\begin{figure}
 \includegraphics[width = \columnwidth] {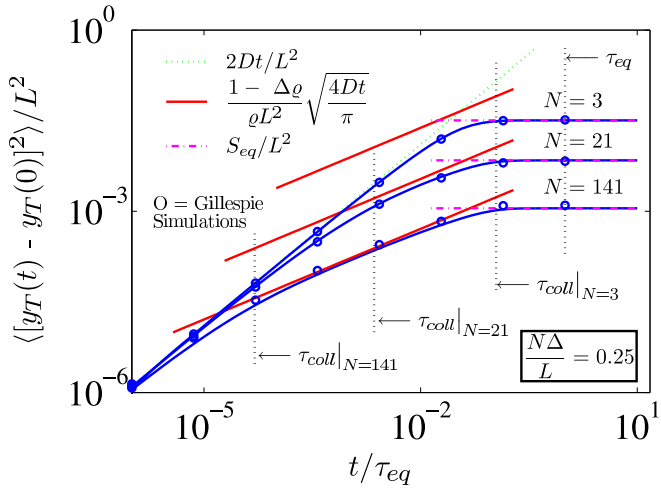}
 \caption{(Color online) Mean square displacement for a tagged
   particle placed in the middle of the ensemble where $N =\{3,21,141
   \}$. Numerical calculations of ${\cal S}(t)$ based on
   Eq.~(\ref{eq:PDF_tagged2}) are indicated by thick blue
   lines. Straight lines show the approximated analytic results for
   regimes $(i)-(iii)$, see text. Results from Gillespie simulations are
   denoted by $(\circ)$ with errors $1-{\cal S}_{\rm
     Gillespie}(t)/{\cal S}(t) < 0.02$.}
 \label{fig:sigma_vs_t}
\end{figure}

{\it Three dynamical regimes. - } Figure~\ref{fig:sigma_vs_t}
depicts a numerical calculation of the mean square displacement of a
tagged particle located in the middle of the ensemble. The solid
blue lines were obtained from numerical integration (trapezoidal
method) of
${\cal S}(t) =
\int_{-L/2+\Delta(N_L+1/2)}^{L/2-\Delta(N_R+1/2)}dy_{\cal T} \,
                    (y_{\cal T}-y_{{\cal T},0})^2
                    \rho_{\cal T}$,
using Eq.~(\ref{eq:PDF_tagged2}) for $N = \{3,21,141\}$. The behavior
seen in Fig.~\ref{fig:sigma_vs_t} illustrates the existence of three
distinct regimes $(i)$-$(iii)$, which become more pronounced as $N$
 increases.


In order to attain a deeper understanding of how regimes $(i)$-$(iii)$
emerge, $\rho_{{\cal}T }$ [Eq.~(\ref{eq:PDF_tagged2})] was analyzed
for large $N$, keeping $L$ finite. A saddle-point approximation of the
integral representation of $\Phi(a,b,c;\xi)$
[Eq. (\ref{eq:JacobiPol})] proved unsuitable since it does not hold
for all $\xi\in[0,1]$ ($i.e$ all times).  However, making use of
asymptotic forms of the Jacobi polynomial, derived in
\cite{Elliott_Oliver}, we obtained a large $N$-expansion of
$\Phi(a,b,c;\xi)$ valid for all $\xi$
  \footnote{
  $
  \Phi(a,b,c;\xi)|_{N\gg 1} \approx (N_R-b)!(N_L-a)! / [N-1-(a+b+c)]!
   \left(2/[N-(a+b+c)]\right)^c \xi^{c/2-1/4}
  (1-\xi)^{(N-(a+b+c))/2}
  \zeta^{1/2} I_c[(N-(a+b+c)) \zeta ]
  \{ 1-{\cal O}(N^{-1}) \},
  $
  where $I_c(z)$ is the modified Bessel function of the second kind and
  $\zeta=\log [(1+\sqrt{\xi})/(1-\sqrt{\xi})]/2$.
  The regimes are given by $(N\gg 1)$:
  $(i)$   $N\zeta\ll 1$, $\zeta\ll 1$;
  $(ii)$  $N\zeta\gg 1$, $\zeta\ll 1$
  (equivalent to $(y_{\cal T}- y_{{\cal T},0})^2/4Dt\ll 1$);
  $(iii)$ $N\zeta\gg 1$, $\zeta\gg 1$ \cite{AMLI}.},
and asymptotic expressions for $\rho_{{\cal} T}$ in $(i)$-$(iii)$
and crossover times ($\tau_{\rm coll}$ and $\tau_{\rm eq}$), were
deduced:


$(i)$ Short times ($t\ll \tau_{\rm coll}$): For short times very few
particle (wall) collisions have yet occurred and the particles are
(to a good approximation) diffusing independently of each other. In
this limit, $\rho_{{\cal}T }$
%
%
is Gaussian
$ \rho_{\cal T} = (4\pi Dt)^{-1/2}
       \exp[-(y_{\cal T}-y_{{\cal T},0})^2/(4Dt)]$,
with mean square displacement ${\cal S}(t)=2Dt$, which is in agreement
with the numerical integration of Eq.~(\ref{eq:PDF_tagged2}), see
Fig.~\ref{fig:sigma_vs_t}.


$(ii)$ Intermediate times ($\tau_{\rm coll} \ll t \ll \tau_{\rm eq}$):
The dynamics in this regime is dominated by particle collisions, and
single-file behavior is observed:
${\cal S}(t)\propto t^{1/2}$ (Fig.~\ref{fig:sigma_vs_t}). The PDF in
this regime (for a particle located not too close to the edge) is
found to be
\begin{eqnarray}\label{eq:Gaussian_SFD}
  \rho_{\cal T}  =
     \frac{1}{\sqrt{2\pi}} \hspace{-0.1cm}
          \left(
           \frac{1}{\frac{4Dt}{\pi} \left(\frac{1- \varrho\Delta}{ \varrho}\right)^2}
          \right)^{1/4}
     \hspace{-0.45cm}
             \exp\left(
               -\frac{(y_{\cal T} - y_{{\cal T} 0})^2}
               {2\sqrt{\frac{4Dt}{\pi}
               \left(\frac{1- \varrho\Delta}{ \varrho}\right)^2}}
              \right) \ \
\end{eqnarray}
which is a Gaussian with a concentration dependent mean square
displacement ${\cal S}(t) =[(1- \varrho\Delta) / \varrho]
\sqrt{4Dt/\pi}$. Thus, the simple rescaling $\varrho\rightarrow
\varrho/(1-\varrho \Delta)$ takes us from the point-particle case
\cite{MK,TEH,JALA} to the finite particle case.



$(iii)$ Large times ($t\gg \tau_{\rm eq}$): For large times,
$\rho_{\cal T}$ reaches equilibrium [Fig. \ref{fig:Densty_vs_t}(b)
contains examples], and ${\cal S}(t)$ is constant for $t > \tau_{\rm
eq}$ (Fig. \ref{fig:sigma_vs_t}). The equilibrium density $\rho_{ {\rm
eq}, {\cal T}}$ is found using $\lim_{t \rightarrow \infty}\xi = 1$,
leading to $\lim_{\xi \rightarrow 1}\Phi = 1$
\footnote{The function $\Phi(a,b,c;\xi)$ can be expressed in terms of
the Gauss hypergeometric function $\Phi(a,b,c;\xi) = { }_2F_1[-N_L+a,
-N_R+b, -(N_L+N_R)+a+b+c;1-\xi]$ for which $\lim_{\xi \rightarrow
1}\Phi(a,b,c;\xi) = 1$.},
and a large $t$ expansion of Eq.~(\ref{eq:psi_int}):
\begin{eqnarray} \label{eq:PDF_eq2}
 \rho_{{\rm eq},{\cal T}} &=& \frac{1}{(L-N\Delta)^N}
  \frac{(N_L+N_R+1)!}{N_L!\,N_R!} \nonumber \\
 &&
  \times
  \left(\frac{L}{2}+y_{\cal T}-\Delta(1/2+N_L) \right)^{N_L} \nonumber \\
&&
  \times
  \left(\frac{L}{2}-y_{\cal T}-\Delta(1/2+N_R) \right)^{N_R}.
\end{eqnarray}
Notably, Eq.~(\ref{eq:PDF_eq2}) is recovered by direct integration of
Eq.~(\ref{eq:PDF_eq1}), and also from simple entropy arguments
\cite{AMLI}. The mean square displacement ${\cal S}(t\rightarrow
\infty)\equiv {\cal S}_{\rm eq} $ for the case $N_L=N_R$ reads
\begin{equation} \label{eq:MSD_eq}
 {\cal S}_{\rm eq} =
 \left(\frac{1}{4}\right)^{N_R+1} \left(\frac{L-N\Delta}{2}\right)^2
 \frac{\Gamma(1/2) \Gamma(2(N_R+1))}{\Gamma (N_R+1)\Gamma(N_R + 5/2) } ,
\end{equation}
where $\Gamma(z)$ is the gamma function.
%

%
%

{\it Conclusions.- } We have found an exact solution to a
non-equilibrium many-body statistical mechanics problem involving
finite-sized particles diffusing in a finite system. The analysis
showed, for the first time, the existences of three distinctly
different dynamical regimes for which exact analytical expressions of
the PDF were found, using a non-standard asymptotic technique. The
results showed excellent agreement with Gillespie simulations. 

The motion of tagged particles is sensitive to environmental
conditions ($e.g.$ concentration, diffusion constant and system size),
suggesting that fluorescently tagged particles can function as probes
or sensors at the nanoscale.

%
%

We thank Owe Orwar, Bob Silbey, Mehran Kardar, Ophir Flomenbom and
Michael Lomholt for valuable discussions and comments.
T.A. acknowledges the support from the Knut and Alice Wallenberg
Foundation.


\end{document}